# Integrated electronic controller for dynamic self-configuration of photonic circuits

Emanuele Sacchi[1,*], Francesco Zanetto[1], Andrés Ivan Martinez[1], SeyedMohammad SeyedinNavadeh[1], Francesco Morichetti[1], Andrea Melloni[1], Marco Sampietro[1] and Giorgio Ferrari[2]

[1]Department of Electronics, Information and Bioengineering, Politecnico di Milano, piazza Leonardo da Vinci 32, Milano, 20133, Italy.

[2]Department of Physics, Politecnico di Milano, piazza Leonardo da Vinci 32, Milano, 20133, Italy.

* Corresponding author: emanuele.sacchi@polimi.it

## Abstract

Reconfigurable photonic integrated circuits (PICs) can implement arbitrary operations and signal processing functionalities directly in the optical domain. Run-time configuration of these circuits requires an electronic control layer to adjust the working point of their building elements and make them compensate for thermal drifts or degradations of the input signal. As the advancement of photonic foundries enables the fabrication of chips of increasing complexity, developing scalable electronic controllers becomes crucial for the operation of complex PICs. In this paper, we present an electronic application-specific integrated circuit (ASIC) designed for reconfiguration of PICs featuring numerous tuneable elements. Each channel of the ASIC controller independently addresses one optical component of the PIC, as multiple parallel local feedback loops are operated to achieve full control. The proposed design is validated through real-time reconfiguration of a 16-channel silicon photonics adaptive beam coupler. Results demonstrate automatic coupling of an arbitrary input beam to a single-mode waveguide, dynamic compensation of beam wavefront distortions and successful transmission of a 50 Gbit/s signal through an optical free-space link. The low power consumption and compactness of the electronic chip provide a scalable paradigm that can be seamlessly extended to larger photonic architectures.

## Introduction

There has been growing interest and need for general-purpose photonic integrated circuits (PICs) that can be configured to perform all-optical signal processing of light beams [1]. Because of their flexibility, several applications have been proposed for these devices, including reconfigurable filters [2], microwave photonics and beam forming networks [3], separation of mixed guided modes [4], vector-matrix multiplication [5], quantum information processing [6], and neural networks [7]. Reconfigurable and programmable PICs usually consist of matrices of integrated photonic devices, such as Mach-Zehnder Interferometers (MZIs) [8], whose working point can be set on demand to perform dynamic spatial light routing and define the overall photonic functionality [9]. A dedicated electronic control layer is thus needed to enable run-time configuration of the PIC, ideally without requiring any prior calibration. This latter requirement can be achieved by closing around each photonic device a feedback control loop, which assesses the working point in real-time and counteracts unwanted variations from the target. However, as PICs increase in complexity, eventually featuring hundreds of optical devices, the scalability of the control hardware becomes a critical issue. Indeed, currently available discrete components-based electronic boards [10] do not represent a viable option due to their bulkiness and power consumption, which make them suitable for laboratory demonstrations but unpractical in commercial applications.

Application-specific integrated circuits (ASICs) offer a promising alternative, as their intrinsic scalability and modularity allow to drive each photonic device independently, consuming only a fraction of the power and area required by an electronic board. Solutions have already been proposed to implement integrated local feedback loops that stabilize the working point of PICs containing few devices against thermal drifts [11, 12]; what is still missing is a CMOS architecture that seamlessly extends this approach to larger optical processors, featuring tens of devices, and enables dynamic reconfiguration of PICs, thus opening applications requiring more than the sheer compensation of slow time-varying drifts. To some extent, the "application-specific" paradigm has to be moved from photonic to electronic circuits [1], with little to no loss in terms of programmability of the optical processor, as conceptually shown in Figure 1a. A secondary motivation behind the adoption of ASIC controllers can be found considering the fabrication flow involved in integrated optics, which largely relies on long-established processes already developed for CMOS electronic integrated circuits [13]: it is straightforward to think of a chiplet-like electronic-photonic co-design, where the same package contains both ASICs and PICs, which can even be arranged in a flip-chip architecture [14]. Monolithic integration of electrical and optical functionalities on a single silicon die has also been proposed [15, 16], although it still comes with trade-offs between performance and cost.

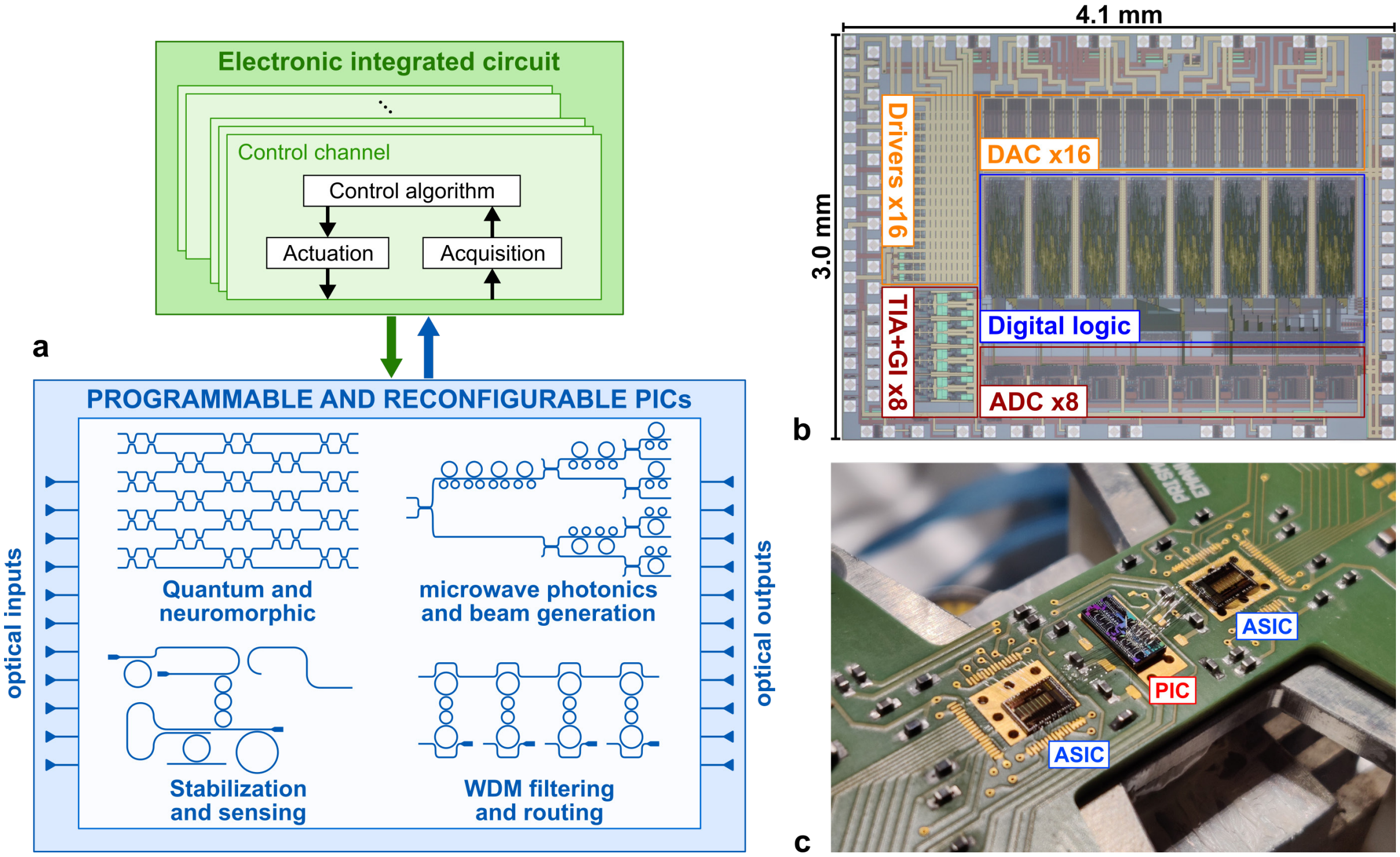

**Fig. 1 Integrated electronic controller for reconfigurable photonic circuits.** a) Schematic view of some programmable photonic circuits that would greatly benefit from connection to a dedicated electronic control layer, capable of reconfiguring the optical functionality at run-time. b) Microscope photograph of the custom electronic ASIC for real-time control of programmable PICs described in this work, highlighting its size and main internal sections. c) Connection of two ASICs to a programmable photonic circuit, demonstrating the compact size of the complete assembly that ensures scalability to large-scale optical architectures.

To fill this missing gap, in this work we present an 8-channel electronic application-specific integrated circuit (ASIC) designed to perform real-time configuration of programmable photonic chips, requiring no prior calibration to set and maintain the optical functionality effectively. To validate the approach, a pair of ASICs were used to control a 16-channel self-aligning universal beam coupler [17, 18] fabricated in Silicon Photonics technology. Experiments were conducted to assess the ability of the CMOS controller to automatically determine the optimal set of working points of each optical device that maximizes the coupling efficiency between the programmable PIC and the input beam. Time-varying effects affecting the optical link, including dynamic perturbations of the wavefront, were successfully compensated, eventually enabling a 25 Gbaud communication through an indoor free-space setup, with a power consumption lower than 10 mW per channel and an overall area occupation of 12 mm$^2$. The proposed ASIC thus represents a key element for supporting scalable and reliable use of photonic integrated circuits in all those applications where static calibrations and lookup tables are not a sufficient solution, such as all-optical signal processors, general purpose programmable PICs and adaptive and reconfigurable devices.

## Results

### Architecture of the integrated CMOS controller

The 8-channel CMOS electronic integrated circuit has been specifically designed for dynamically controlling programmable optical circuits implemented in Silicon Photonics technology. The ASIC has been conceived to operate PICs whose functionality is defined by completely switching on/off optical interconnections, implemented with Mach-Zehnder interferometers. When dealing with MZIs, each channel of the controller simultaneously drives two thermo-optical actuators; however, the chip can also be effectively employed to control MRR-based circuits, where only one actuator is needed, as well as combinations of MZI and MRR and any other photonic circuit. Figure 1b shows a microscope photograph of the CMOS controller. The ASIC, built in AMS CMOS 0.35 μm technology, occupies an active silicon area of ≈12 mm$^2$, comparable to that of the photonic chip. The power dissipation of each electronic channel, working with 3.3 V power supply, is ≈10 mW, less than what a single thermal actuator requires to achieve a phase shift larger than $2\pi$, in the order of ≈30 mW (measurements are provided in Supplementary Section 4). Given the compact size of the electronic chip, multiple ASICs can be connected to a single PIC whenever more than 8 channels have to be managed in parallel. Figure 1c shows the connection between two ASICs and a PIC (described in detail in the following section), made through direct wire bonding to minimize the effect of stray capacitance, maximize the readout accuracy and minimize the assembly size. A custom printed circuit board (PCB) has been designed to host the chips during the experimental validation, route the electrical signals necessary for their operation and communicate with a personal computer for configuration and monitoring of the system.

The architecture of the electronic control channel is shown in Figure 2a. The optical power impinging on one output branch of the photonic device is read by an integrated photodiode (PD). In this work, we consider germanium PDs that are commonly provided by foundry process design kits, but other kinds of PDs or monitor photodetector [19, 20] can be used as well by properly adapting the circuit. The photogenerated current is fed to the analog frontend of the ASIC for amplification through a transimpedance amplifier, lowpass filtered with a gated integrator and digitized by a 10-bit analog-to-digital converter (ADC) working with sampling rate $f_S$ = 100 kSamples/s. An input optical power dynamic range of 50 dB has been considered during the sizing of the analog acquisition chain to correctly detect the photogenerated current (≈30 dB to account for the MZI/MRR rejection ratio, plus an additional margin of 20 dB to adapt to possible coupling losses and different optical power levels). The gain of the front-end stage is automatically adjusted by a 6-steps digital logic according to the absolute optical power impinging on the photodetector, in order to always match the ADC full-scale range to the photocurrent level. Optical signals between 0 dBm and −50 dBm can thus be successfully detected and managed by the ASIC, suitable for realistic applications. This is confirmed by the experimental measurement shown in Figure 2b, which reports the values sampled by the ADC as a function of the input PD current and highlights the operation of the adaptive amplification stage.

The control algorithm to maximize/minimize the optical power at the output of each photonic device, needed to switch on/off the desired optical paths, is implemented at digital level and relies on the dithering technique in combination with integral controllers [21]. This approach allows us to define a feedback strategy that is easily scalable to multiple devices and independent of the average light power in the chip and the setup temperature. More details on the control algorithm are provided in Supplementary Section 1. The dithering extraction and integration processes are performed in the digital domain by two identical chains, for maximum flexibility of the chip. Two chains operating on orthogonal dithering signals are needed when controlling MZIs with two heaters, whereas a single digital circuit is enough for a MRR. Within each chain, a square-wave multiplier downconverts the ADC readout and a digital integrator accumulates the resulting signal, thus simultaneously implementing the integral feedback law needed for controlling the photonic device and precisely defining the control loop bandwidth. The obtained DC signal is used to update the device working point, upon which the square-wave oscillation is superimposed in order to apply the dithering modulation to the actuators. A square root operator, whose details are described in Supplementary Section 4, is used to compress the generated digital word and linearize the feedback loop, by compensating the quadratic relation between the heater voltage and its dissipated power. This ensures the same control loop response regardless of the actuators DC bias [14]. The output digital word is finally compared with two programmable thresholds, that trigger the reset of the accumulator at midscale in order not to saturate the driving circuit when the generated voltage gets too close to the power supply rails. The output of the digital logic is fed to a mixed-signal actuation circuit, made of a pair of 12-bit digital-to-analog converters (DACs) and high-current analog drivers. The drivers are designed to operate the integrated thermal phase shifters, which have a nominal resistance of $R_H \approx$ 400 Ω, within a 0-6 V range, corresponding to a maximum phase shift of ≈4$\pi$. In this way, the control loop is always able to find a minimum in the device transfer function regardless of the algorithm starting condition. The resolution of the DACs has been chosen to achieve an accuracy of ≈1.5 mV in the generation of the actuators voltage, enough to perform an effective control action. Figure 2c shows the measured current provided by the driving stage to the actuators, and the corresponding heater dissipated power, confirming the effective linearization performed by the square root operator and the sufficient power capability of the circuit.

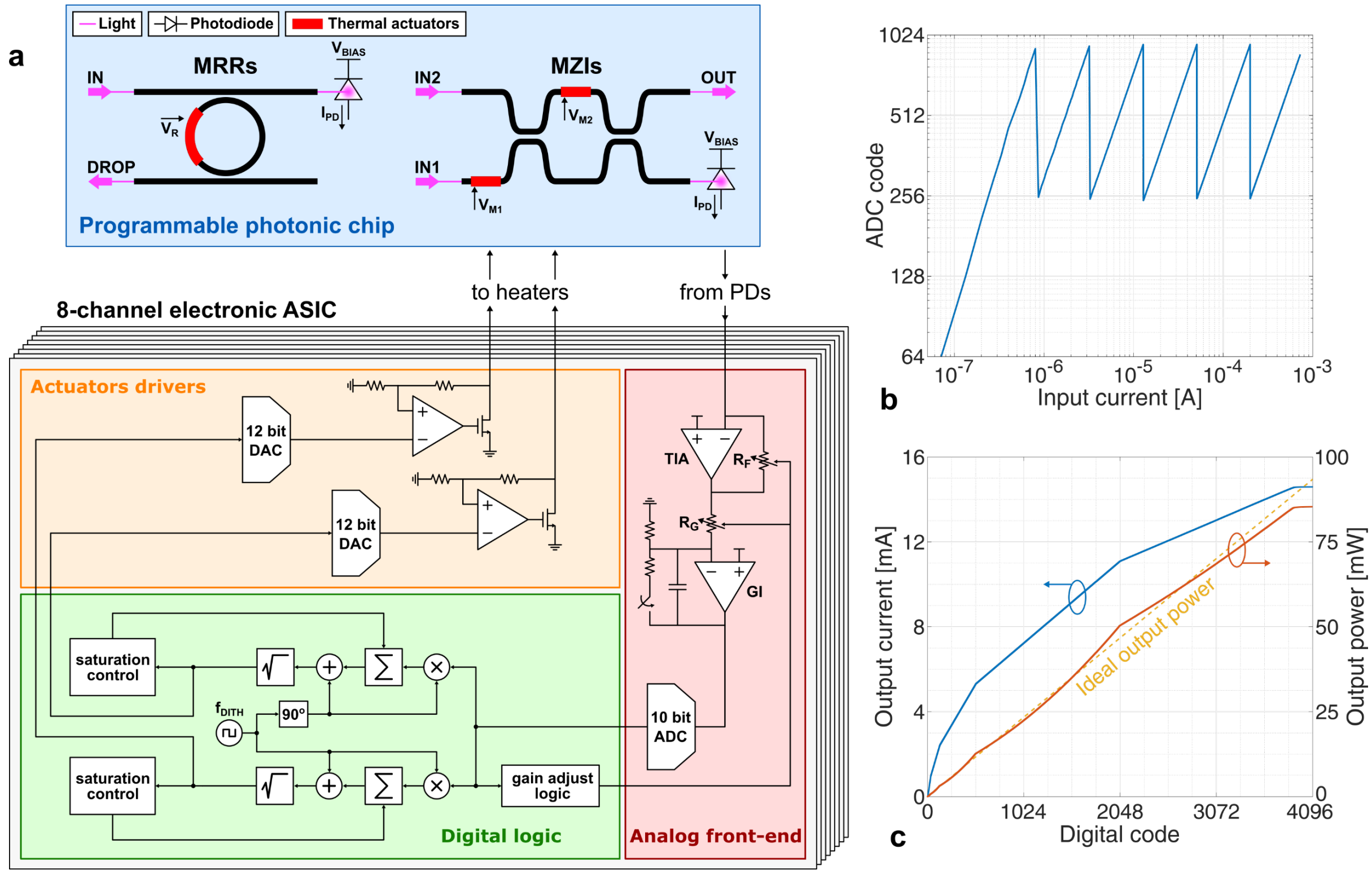

**Fig. 2 Electronic chip design and characterization.** a) Schematic view of the ASIC architecture for dynamic control of programmable photonic circuits. The ASIC features 8 parallel channels. b) Measured ADC code for a wide range of input currents, that can be correctly detected thanks to the adaptive amplification mechanism. c) Characterization of the actuator driver circuit, validating the linear control of the actuators power achieved through square root compression.

## Integrated 16-channel self-aligning beam coupler

The designed ASICs have been used to dynamically control a 16-channel silicon photonics integrated self-aligning universal beam coupler (UBC). UBCs are a class of reconfigurable PICs that can receive an arbitrary monochromatic input beam and be configured in order to couple it to a single-mode waveguide [17, 18]. They can be implemented as a combination of 2×2 interferometers, such as Mach-Zehnder interferometers (MZIs), arranged in a mesh configuration [10, 22]. By making each interferometer tunable with proper on-chip actuators, an adaptive UBC can be dynamically reconfigured to work as a coherent intensity adder and constantly maximize the input-output coupling efficiency even when the input beam characteristics change. The reconfigurability of UBCs makes them suitable to track and mitigate the effect of dynamic phenomena, such as time-varying angular misalignments (i.e. direction of arrival) between an input beam and the photonic chip and/or wave-front distortions caused by propagation through aberrators, scattering media, multimode or multicore fibers/waveguides or turbulent environments, as in the case of atmospheric turbulence in free-space optical (FSO) links [23].

A schematic of the designed integrated photonic circuit and its top-view microscope photograph are shown in Figures 3a and 3b. The circuit has been manufactured by a commercial silicon photonic foundry (AMF, Singapore [24]) and has an area of 3.2 mm × 1.3 mm. It consists of 15 thermally-tunable balanced MZIs arranged in a 4-stage 16×1 binary mesh. Compared to other mesh topologies [17, 22], this arrangement equalizes the optical path from all inputs to the output and minimizes the number of MZI cascaded stages, thus resulting in a faster configuration time for the receiver. At the input (left-hand side in Figure 3a), the MZI mesh is connected to an integrated 16-element optical antenna array (OAA). The antennas are realized by using surface grating couplers (GCs) designed to operate on transverse-electric (TE) polarized light. The 16 GCs are arranged in two concentric rings with a radius of 60 µm (inner ring, 7 GCs) and 180 µm (outer ring, 8 GCs), respectively, and a central GC, as shown in Figure 3b. This OAA configuration is optimized to receive optical beams with a linear polarization and circular symmetry, but other OAA topologies can be designed for other classes of beams [25]. Waveguide stubs are inserted between the OAA and the MZI mesh to match the length of each input optical path and minimize the wavelength dependence of the UBC [22] (see Methods for further details). At the output of the MZI mesh, 15 waveguides are coupled to integrated photodetectors, while the 16th waveguide (optical output) is terminated with a GC for coupling with an optical fiber.

Two ASICs have been used to control the UBC. To demonstrate the possibility of configuring the photonic circuit starting from any initial condition, the heater voltages have been initially set to random values, then the 16 control loops have been simultaneously activated. Figure 3c shows the measured PD currents during the configuration transient of the photonic circuit. All the photocurrents are minimized by the ASICs, thus steering all the input light towards the optical output of the chip. The configuration of the first stage of the receiver takes place in around 1.5 ms, while the following layers are characterized by a slightly longer convergence time. In fact, the $N^{th}$ stage of the mesh cannot be completely tuned before all the previous $N-1$ stages have been set. This is visible in the figure, which highlights the sequential tuning of the circuit. Even with 4 cascaded stages, the electronic ASICs can configure the whole MZI architecture in just 10 ms, as shown by the transient of the optical output power in the inset of the figure.

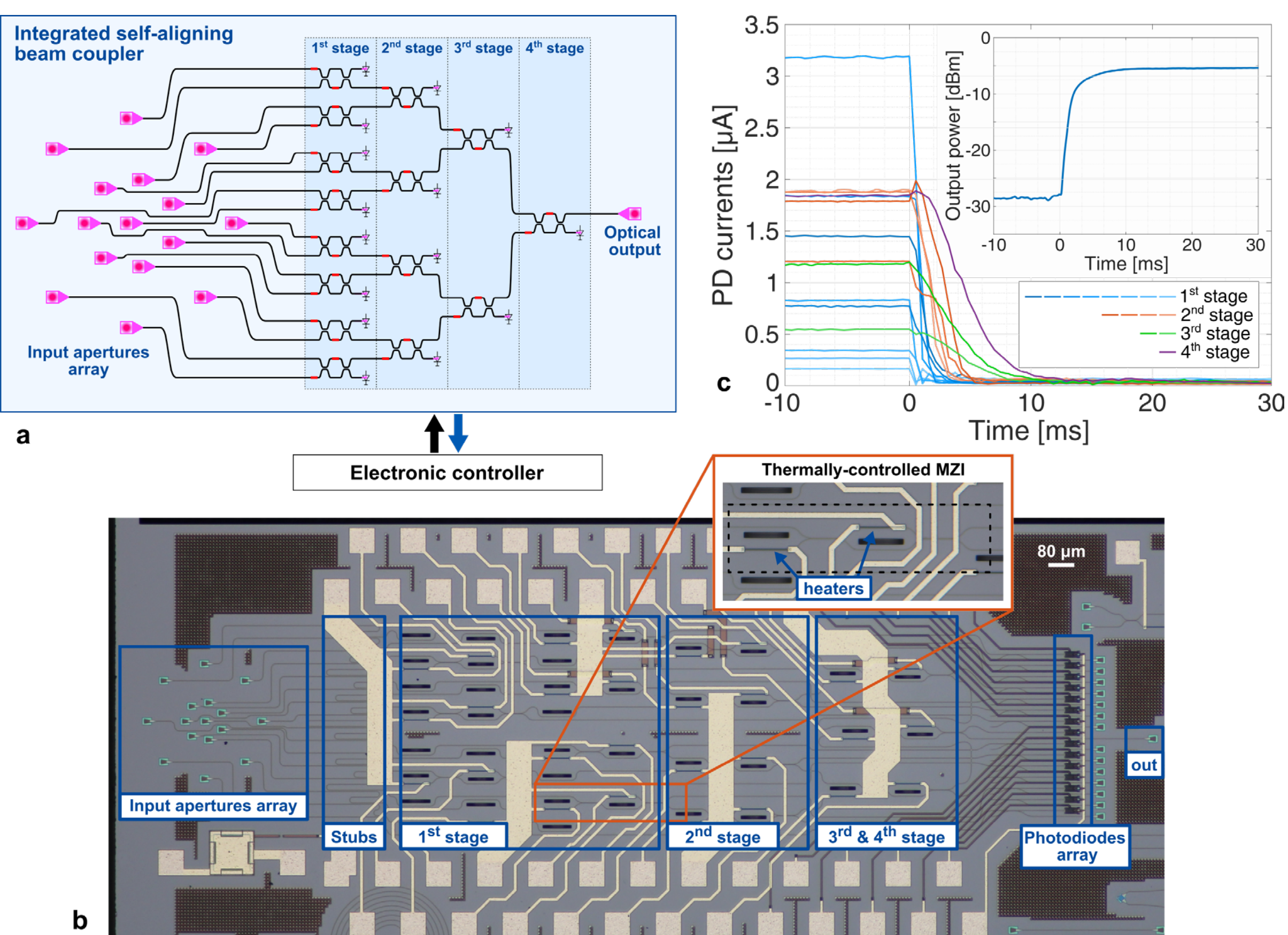

**Fig. 3 Silicon photonics 16-channel self-aligning universal beam coupler.** a) Schematic view and b) microscope photograph of the self-configuring optical beam coupler, made of a binary-tree mesh of 15 MZIs. A detail of a thermally-tunable MZI, featuring two actuators to completely steer the input light to one of the device outputs, is also reported. c) Time transient of the PDs photocurrents when configuring the optical circuit, showing correct minimization when the ASICs are activated after 10 ms. The inset shows the evolution of the chip output power, which is correspondingly maximized in around 10 ms.

## Dynamic correction of wavefront distortions

The performance of the ASIC control was assessed by testing the UBC as a dynamic tracker of wavefront distortions of free space optical beams. In the first experiment (see Figure 4a), static distortions with a well-controllable profile were intentionally introduced by means of a spatial light modulator (SLM) placed along the propagation path of the beam (see Methods and Supplementary Section 2 for details on the experimental setup). The SLM was used to generate 15 synthetic phase masks (see Supplementary Section 3) and an infra-red camera was used to capture the perturbation effects on the optical beam. As reported in Figure 4b, the Gaussian profile of the beam before the SLM (left panel) is significantly altered after reflection on the SLM, with a significant change of shape, size, and centroid of the beam. The beam is then coupled with the OAA of the PIC through a collimating lens. The performance of the ASIC-controlled PIC was assessed by measuring the optical power at the output port of the MZI mesh when the ASIC control is switched off (all the heaters are held at a constant voltage) and when the feedback control is activated to find the best configuration of the mesh and maximize the output power. Figure 4c shows the optical output power for different phase screens of the SLM (data are normalized to the output power received when SLM is off and no perturbation is introduced). When the control is turned off, the average received power is −5.7 dB with a standard deviation of 2.9 dB. Instead, when the UBC is automatically reconfigured after changing the SLM mask, the average power improves to −0.8 dB and the standard deviation reduces to 0.6 dB, thus demonstrating the effectiveness of the adaptive ASIC controller. The slight residual fading is due to the shift of the centroid of the beam causing variations of the total power impinging on the GCs of the OAA under different perturbation profiles. This can be improved with a larger OAA and with a higher fill factor, obtained by employing more GCs, an array of lenslets or a photonic lantern device [22].

To test the ASIC controller in dynamic conditions with fast wavefront distortions of the optical beam, the SLM was replaced with a heat gun introducing turbulence in the free space path (Figure 5a). Figure 5b shows the measured power at the optical output of the PIC when the control ASICs are activated (blue curve) /deactivated (orange curve) (data are normalized to the output power received when the PIC is configured for an unperturbed Gaussian beam, obtained with the heat gun turned off). When the ASICs are on, the UBC compensates for the wavefront distortion and the detected optical power at the output has a mean value of −0.17 dB, a standard deviation of 0.1 dB and a maximum fading of less than 0.5 dB. After 8 s, the control loops are paused, by holding the heater driving voltages at fixed values. In this condition, the PIC is on average well configured to receive the incoming beam, however the random phase front fluctuations result in a degradation of the received power: a mean value of −1.15 dB and a standard deviation of 0.5 dB are observed, with a maximum fading higher than 3 dB. The same is clearly visible in Figure 5c, which compares the probability

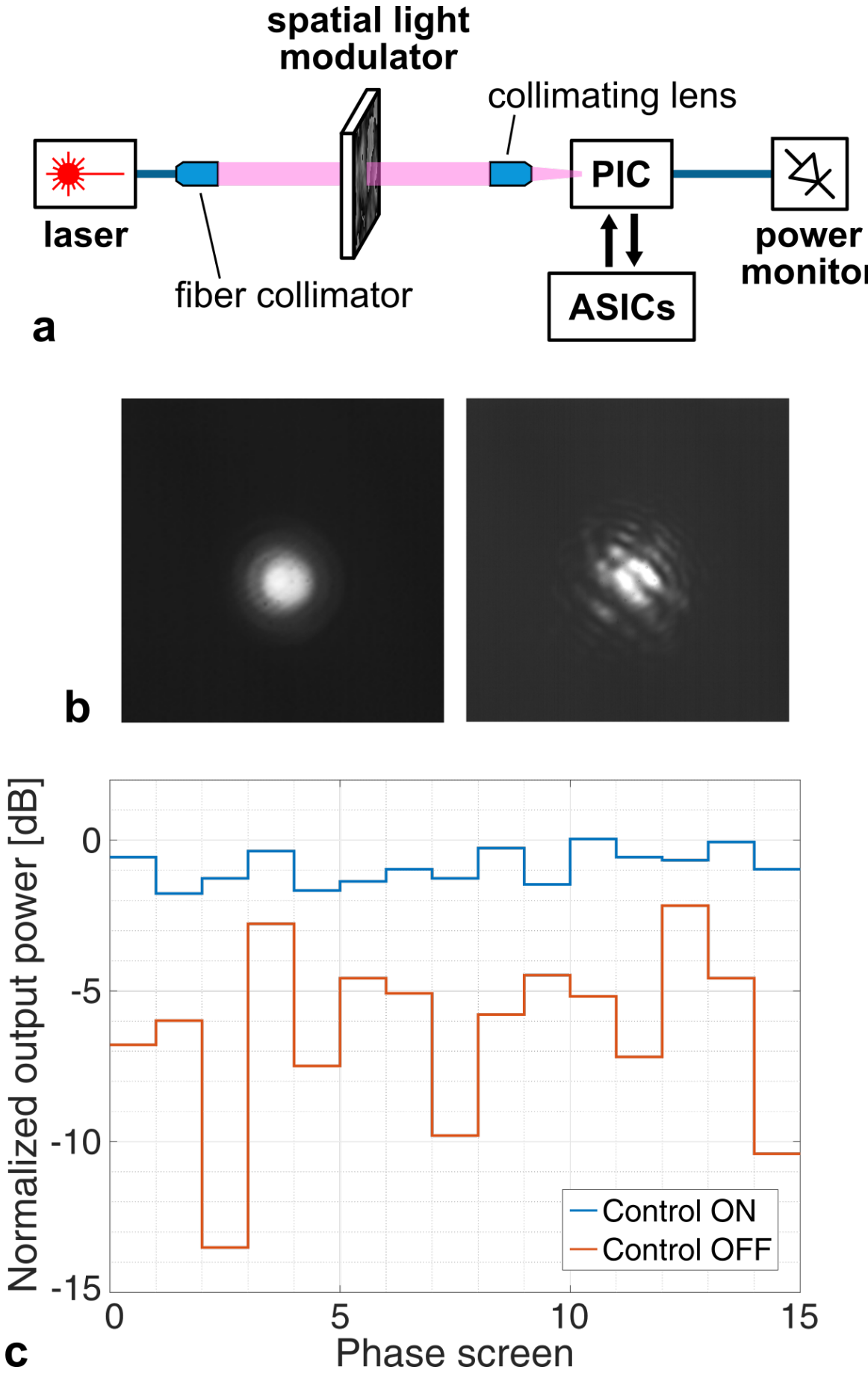

**Fig. 4 Experimental validation.** a) Schematic view of the setup employed to introduce a static perturbation in the wavefront of a free-space beam. b) Photograph, captured with an infrared camera, of the optical beam impinging on the beam coupler, when the SLM is off (left) and when it is used to perturb the free-space beam (right). c) Optical power at the output port of the beam coupler for different SLM phase screens, showing its ability to perform real-time wavefront correction and beam reconstruction.

219 density function (PDF) of the received optical power in the two situations. Results confirm the
220 advantage of dynamic and real-time PIC reconfiguration, which results in a narrower distribution
221 concentrated at higher power levels: 90% of the samples are above the threshold of −0.3 dB when
222 the controller is tracking the injected turbulence, whereas the same parameter worsens to −1.83 dB
223 when no control is exerted (and only ≈1% of the samples stay above −0.3 dB). Figure 5d shows the
224 frequency spectrum of the optical signal at the output waveguide of the UBC. When the control loops
225 are off (orange curve), spectral components extending up to about 300 Hz are observed, which are
226 related to the harmonics of the perturbation generated by the heat gun. The spectrum acquired after
227 the activation of the ASIC controllers (blue curve) confirms that the real-time reconfiguration of the
228 PIC significantly reduces the perturbation effect on the output up to around 300 Hz. This
229 demonstrates that the tracking time of the ASIC-controlled PIC is faster than 10 ms. Notably, this
230 response time is faster than the dynamics of atmospheric turbulence [26], meaning that our
231 approach can be effectively used to compensate for wavefront distortion in real free space links for
232 communication, sensing and ranging applications.

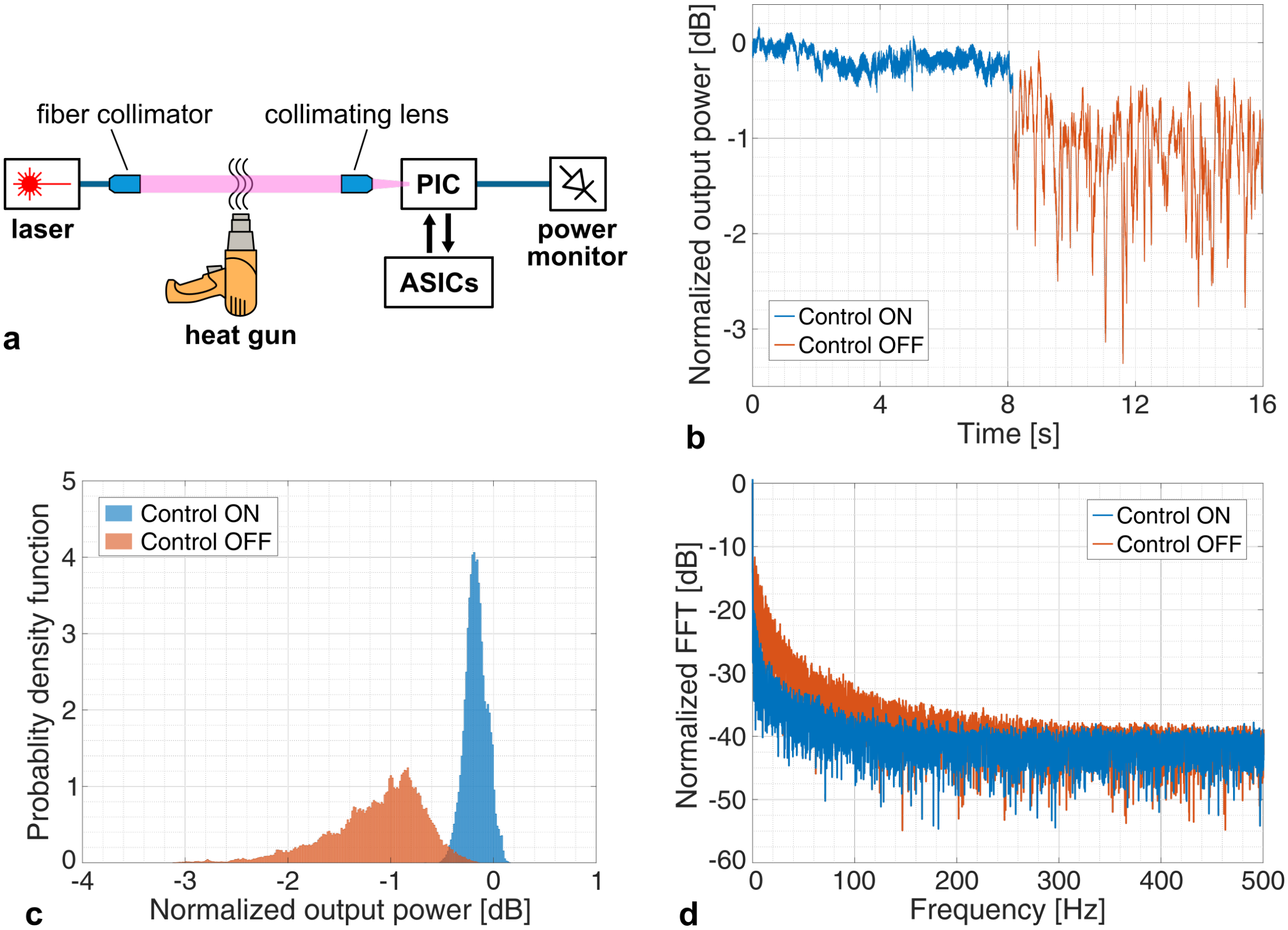

**Fig. 5 Real-time compensation of dynamic perturbations.** a) Experimental setup employed to introduce a dynamic perturbation in the free-space propagation of the beam. b) Temporal evolution, c) probability density function and d) frequency spectrum of the received optical power when the ASICs are active (blue curves) or disabled (orange curves), demonstrating correct real-time compensation of the beam-front distortion performed by the electronically-controlled photonic chip.

## ASIC-assisted adaptive free-space optical receiver

The ASIC-assisted UBC was employed to receive a high-data-rate free-space optical signal corrupted by wavefront distortions. To this aim, the laser beam was modulated in intensity at a symbol rate of 25 Gbaud with a commercial modulator and transmitted through a free-space optical link. As in previous experiments, at the receiver side the optical beam is coupled with the OAA of the UBC, whose optical output is coupled with a single-mode optical fiber and monitored with a real-time optical oscilloscope to assess the signal quality. No signal equalization has been performed. The results of the data transmission experiment are shown in Figure 6a, for the case of 25 Gbit/s non-return-to-zero on-off keying modulation. When the working point of the photonic processor is not optimized by the ASICs (bottom panel), the input beams sampled by the OAA are combined along the MZI mesh with random phase relations, resulting at the chip output in a degraded signal with a completely closed eye diagram. In contrast, a fully open eye diagram with a quality factor of 6.12 is recorded when the ASIC controllers are activated. Notably, the control of the coherent adder performed by the ASIC controller is independent of the modulation format and data rate of the transmitted signal. Figure 6b shows the eye diagrams of a 50 Gbit/s 4-level pulse-amplitude modulation (PAM4) with (upper panel) and without (lower panel) adaptive PIC configuration, confirming a clear improvement of the quality of the received signal when the ASIC controller is activated.

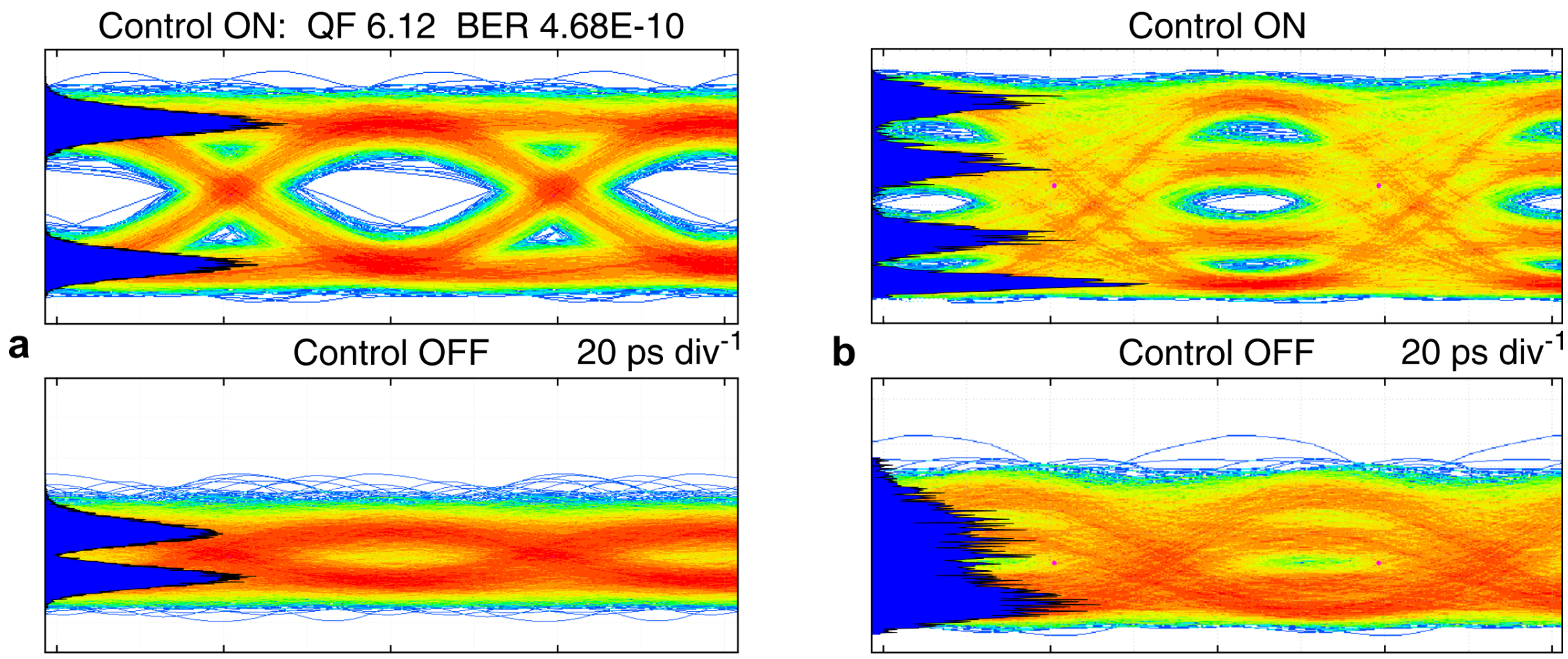

**Fig. 6 High-speed free-space optical transmission.** Eye diagrams of the received optical signal when transmitting a free-space beam modulated at 25 Gbaud both with a) OOK and b) PAM4 modulations, showing the transmission improvement when the adaptive receiver is dynamically operated.

## Discussion

Our results demonstrate dynamic configuration, control and tracking of PICs made of many tuneable elements by using parallel feedback loops integrated in a single multi-channel ASIC controller. The ASIC can automatically set the working point of the photonic chip in real-time and lock the photonic functionality to the desired one. The compact size (similar to a PIC) and low power consumption (the entire electronic channel dissipates less power than a single heater) make the ASIC a pivotal solution for scaling the complexity of programmable PICs to several tens of devices, a level that can be hardly managed with off-the-shelf solutions.

As an example of application, we validated the use of two 8-channel ASICs to dynamically operate a 16-channel thermally tuneable MZI silicon photonics mesh as an adaptive coherent adder. The ASIC controller enables both automated configuration of the programmable PIC and effective compensation of dynamic wavefront distortions of the input beam, with no need for any prior calibration nor knowledge of the transmission channel introducing wavefront distortions. The combination of ASICs and PIC allowed us to automatically couple a free-space beam to a single-mode waveguide even in the presence of dynamic wavefront perturbations, thanks to a control bandwidth around 400 Hz. Successful reconstruction of a 50 Gbit/s PAM-4 modulated signal demonstrated the correct operation of the electronic-assisted UBC, validating the operation of the electronic controller.

The use of electronic-assisted UBCs holds significant potential in FSO communications, imaging and sensing. These systems can be used as self-adaptive receivers to mitigate in real-time the effect of atmospheric turbulence in the link without requiring any digital signal processing at the data rate [23]. In addition, they can be exploited for beam tracking, both in angle and focus, and provide information on the alignment between transmitter and receiver. Other applications can be also envisioned where an UBC can be used as a sensing element, including wave-front analysis, phase-front mapping and reconstruction, imaging through scattering media and obstacle identification [27]. In these situations, the configuration of the on-chip actuators provides the information of interest once the UBC is correctly tuned [28]. Therefore, a mixed-signal control ASIC is advantageous over purely analog solutions, since the heater voltages can be easily monitored digitally through a serial communication interface. Similarly, the possibility of using a UBC to generate structured light, that is, optical beams with well-defined amplitude, phase, and polarization profiles, can be exploited in high-resolution imaging [29] and particle detection and manipulation [30]. In this case, a pre-computed actuators configuration can be uploaded to the ASIC to generate the beam shape of interest, once again thanks to the chip digital read/write interface.

Scalability of the integrated electronic controller is greatly beneficial since it allows operating PICs with a large number of elements, that are required for accurate beam manipulation. Even though the ASIC already provides great scalability in controlling programmable PICs, further improvements are possible. The ASIC developed in this work is suitable for those applications where optical signals need to be completely steered towards a well-defined output, such as UBCs, optical switches and WDM filters. However, several applications require partial light splitting among different optical paths. This is the case, for example, of vector-matrix multiplication circuits [5], mode unscramblers [4] and optical neural networks [31]. In these situations, a different control strategy is needed (one possibility is demonstrated in [32]) requiring an adaptation of the ASIC digital structure to be implemented. In addition, minimally-invasive PDs [20] can be particularly beneficial in these applications, since they can be placed in-line with the optical path without significant penalties. Therefore, a front-end circuit tailored for these devices can also be envisioned. These improvements would make control ASICs versatile and suitable for a wider range of applications and programmable PICs, further boosting its pivotal role.

Technological improvements in the ASIC manufacturing are another possible evolution of this work. A more advanced electronic technology node can be used to fabricate the ASIC, significantly reducing the area of the single control channel and thus allowing the integration of more feedback loops on the same die. This simplifies the assembly and packaging procedure when large-scale PICs need to be operated, improving also the reliability of the whole system. The ultimate goal is to design a control chain with an area occupation similar to a single photonic device, so that the PIC and the ASIC can have the same size. With this approach, the ASIC could be connected in a flip-chip arrangement to the PIC, acting as both an optical element and a mechanical substrate. In this way, the area occupation of the overall system can be minimized. A scaled electronic technology node can also be beneficial in decreasing the ASIC power consumption. First, the dissipation of the digital blocks scales with the transistor size, therefore smaller devices improve the efficiency of the chip. Similarly, scaled transistors enable the design of more advanced actuator driving stages (for example, employing pulse-width modulation techniques [33]), that also contribute in the reduction of the chip power consumption. With these improvements, the ASIC becomes attractive also when non-thermal actuators with high energy efficiency (e.g. devices exploiting materials with high electro-optic coefficients or opto-micromechanical phase shifters) are employed.

# Methods

## ASIC fabrication and characterization

The CMOS chip has been designed in order to read and control the operating point of each photonic device with proper accuracy, considering a maximum rejection of ≈30 dB between the output ports. An optical power impinging on the PD ranging from 0 dBm to −20 dBm has also been considered, therefore the circuit must be able to sense optical power variations as small as −50 dBm. In principle, this would require a complex 16-bit ADC; instead, a simple 10-bit converter has been employed. The required precision has been achieved by automatically adjusting the analog gain of the acquisition chain according to the average impinging optical power: every time it scales by a factor 4, the amplification implemented by the analog channel in Figure 2a is increased accordingly, thus always guaranteeing a rail-to-rail mapping of the given current range. At the same time, when the analog gain doubles, the weight of the ADC samples fed to the digital accumulator is halved, keeping the demodulation process coherent with the decreasing optical power.

Concerning the actuators, DACs with 12-bit resolution have been employed to ensure sufficient precision in controlling the operating point of the heaters, as well as to provide a wide variety of dithering modulation amplitudes, usually in the 5-100 mV range. The DACs have been designed with monotonic behaviour, which guarantees that the loop sign remains unchanged during operations and preserves the control stability. Improved linearity in the control response has been obtained thanks to the digital square-root compression performed before the DACs. The operation is implemented by approximating the exact square-root calculation with a piece-wise linear function, whose slope is halved every time the digital input is increased by a factor of 4. The solution requires fewer resources compared to more precise digital architectures [34] and offers a wider operating range compared to analog solutions [14].

Finally, the design features a pair of shift registers for interfacing the chip with a personal computer: one is used to set the working parameters of the circuit, such as the control bandwidth, and to manually configure the operating point of the PIC, which can be useful when characterizing the optical devices; the other monitors the photogenerated current of each PD, as sampled by the ADC, and the voltage applied to each thermal actuator. These registers allow easy electrical characterization of the ASIC, which is available in Supplementary Section 4.

## Photonic chip design and fabrication

The integrated photonic processor has been designed for operation around the 1550 nm wavelength range and fabricated on a standard 220nm SiP platform (Advanced Micro Foundry, Singapore). The photonic chip size is 3.2 mm × 1.3 mm. All the waveguides of the circuit are single-mode channel waveguides with a propagation loss of about 1 dB/cm. The waveguides connecting the grating couplers to the photonic processor have the same geometrical length to minimize the wavelength dependence of the mesh, so that the circuit can have the widest optical bandwidth. Each GC is about 48 µm-long and 23 µm-wide, with a 24 µm-long taper. The elevation angle of the radiated field is 12°, the azimuth angle is 0°, and the divergence is 5.6°× 9.8°.

The MZIs have been designed with 3 dB directional couplers and two TiN thermal tuners. The heaters are placed on top of the MZI input waveguide and on one of the internal interferometer arms.

This enables the control of the relative phase shift between the optical fields at the MZI input ports and the amplitude split ratio of the interferometer, respectively. The thermal tuners have a power efficiency of 20 mW/$\pi$ and a time response of about 10 μs. The maximum power consumption of the photonic processor is about 400 mW in the worst case, i.e. when a $2\pi$ phase shift needs to be simultaneously applied to all the thermal tuners. Scalability of the system to larger-scale circuits would benefit from the use of non-thermal actuators, exploiting materials with high electro-optic coefficients such as lithium niobate or barium titanate integrated in silicon waveguides, or opto-micromechanical phase shifters. More details on the main photonic building blocks employed in the design of the UBC are reported in Supplementary Section 5.

The wavelength range across which the photonic processors can operate is about 35 nm (from 1535 nm to 1570 nm) [22]. This range is mainly limited by the wavelength-selective response of the GCs and by the wavelength dependence of the 3 dB directional couplers of the MZIs. Such wavelength dependence can be reduced by using optimized designs for broadband GCs [35] and directional couplers [36]. The broadband operation of the UBC thus enables simultaneous reception of multiple WDM channels, providing an effective way for scaling the FSO link capacity.

## Experimental setup

The dynamic control of the ASICs has been experimentally tested on a 3 m-long indoor free-space optical link. A laser with 1550 nm wavelength and 5 dBm output power, amplified by 20 dB with a commercial erbium-doped fiber amplifier, has been used in the experiments. The laser output has been launched into free space with a fiber collimator and then expanded to a diameter of around 5 cm with a set of lenses. Intentional distortions in the beam wavefront have been introduced with a spatial light modulator (SLM) and a thermal gun along the propagation path, thus requiring a continuous update of the UBC working point to enable correct signal reception. The beam has been coupled to the photonic chip with an optical system consisting of a 45°-tilted dielectric mirror and a set of collimating lenses. The coherently recombined signal of the last MZI has been coupled out of the chip with a vertically aligned single-mode fiber and measured with bench-top instruments. The temperature of the assembly has been measured with a thermistor placed next to the photonic chip and the whole setup has been kept at 28 °C with a thermo-electric cooler (TEC). More details on the optical setup are reported in Supplementary Section 2.

**Acknowledgments.** This work was supported by the Italian National Recovery and Resilience Plan (NRRP) of NextGenerationEU, partnership on “Telecommunications of the Future” (Program “RESTART”, Structural Project “Rigoletto,” and Focused Project “HePIC”) under Grant PE00000001. The authors thank Polifab, the nanofabrication facility of Politecnico di Milano, for dicing and wire-bonding the chips.

**Data availability.** All the data supporting the findings of this study are available within this Article and its Supplementary Information. Any additional data are available from the corresponding author upon reasonable request.

**Author contributions.** ES and FZ designed the electronic ASICs, performed the measurements and wrote the original manuscript. AIM and SS designed the photonic chip and performed the optical measurements. AM and FM supervised the optical design and characterization. GF and MS led the research and supervised the whole project. All authors contributed to the revision of the manuscript.

**Competing interests.** The authors declare no competing interests.

# Integrated electronic controller for dynamic self-configuration of photonic circuits

# SUPPLEMENTARY MATERIAL

Emanuele Sacchi[1,*], Francesco Zanetto[1], Andrés Ivan Martinez[1], SeyedMohammad SeyedinNavadeh[1], Francesco Morichetti[1], Andrea Melloni[1], Marco Sampietro[1] and Giorgio Ferrari[2]

[1]Department of Electronics, Information and Bioengineering, Politecnico di Milano, piazza Leonardo da Vinci 32, Milano, 20133, Italy.
[2]Department of Physics, Politecnico di Milano, piazza Leonardo da Vinci 32, Milano, 20133, Italy.

* Corresponding author: emanuele.sacchi@polimi.it

## Supplementary section 1: dynamic self-configuration of universal beam couplers

The proposed control strategy for automatically configuring each photonic device relies on the dithering technique to realize a calibration-free feedback loop that does not require prior knowledge of the device transfer function. Here, the main concepts of this technique are briefly summarized, while more details on its implementation can be found in [37, 38].

Figure S1a shows the working principle of the dithering technique applied to a single thermally tunable Mach-Zehnder interferometer (MZI). A small oscillation is superimposed to the DC voltage of each actuator (H1, H2), making the device transfer function fluctuate around its bias point with an oscillation depth proportional to the transfer function first derivative. By choosing orthogonal (e.g. in-phase/quadrature) dithering signals, the effect of the two MZI actuators can be discriminated with a lock-in demodulation of the signal provided by a single photodetector placed at the drop port [39]. In this way, the MZI transfer function and its partial derivatives with respect to the heaters power can be extracted in real time from the physical system.

The information on the partial derivatives is then exploited to control the MZI and minimize the amount of light detected at the drop port. This is obtained by tuning the DC heater voltages to drive the output dithering signals to zero, as this condition corresponds to a stationary point of the MZI transfer function. With respect to other techniques [40–43], this approach does not require any prior calibration, since the condition of null output dithering oscillation only depends on the MZI transfer function and is not affected by any other parameter of the system, such as instabilities of the input light power or temperature drifts. The DC heater voltage is automatically set by two parallel integral controllers, each fed with one of the two partial derivative signals (Figure S1b). The sign of the integral controller gain is adjusted to provide a negative feedback loop only when the minima of the MZI transfer function are reached, thus avoiding ambiguity on the target locking point.

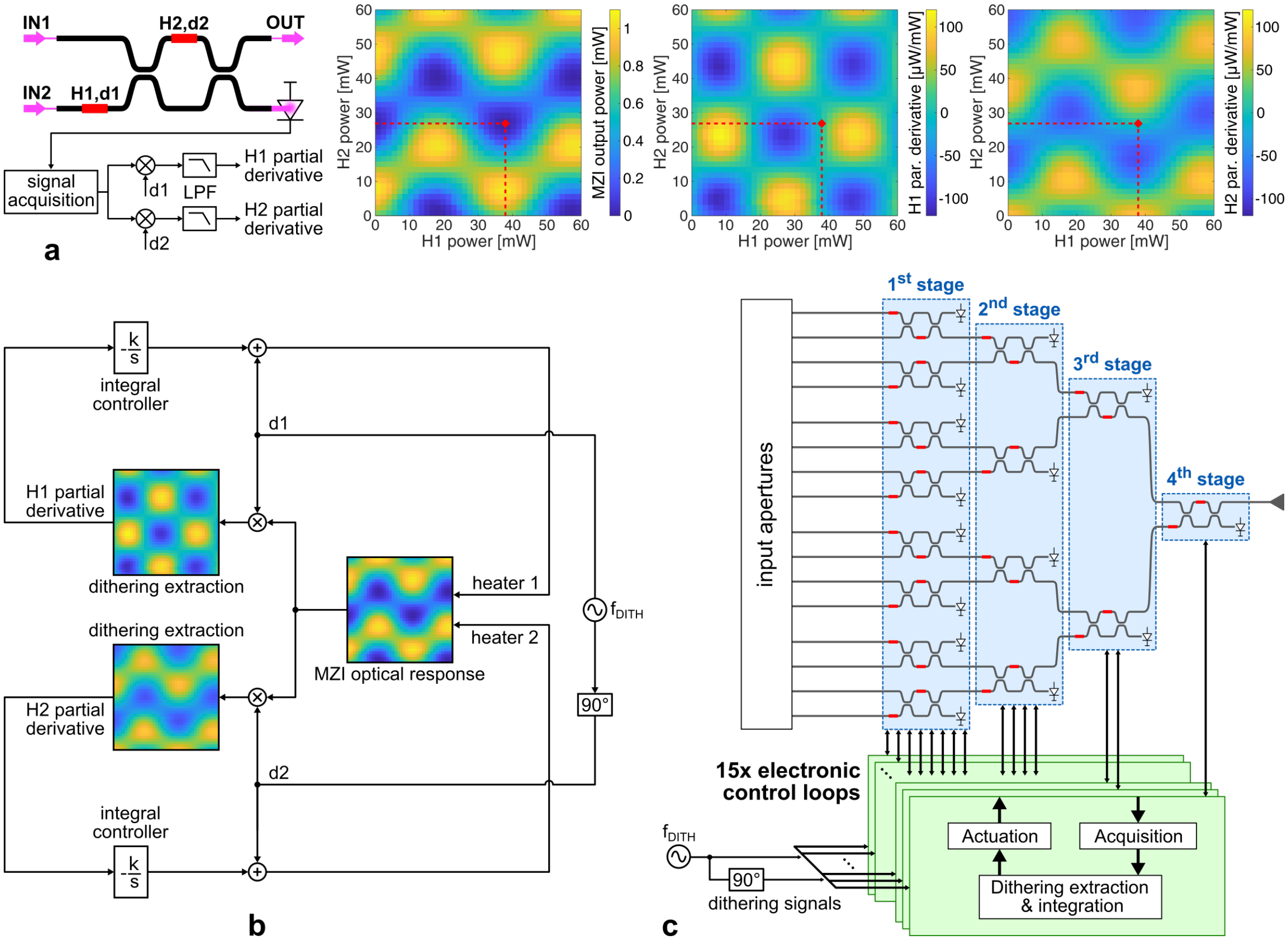

**Fig. S1 Automated control of self-aligning beam couplers.** a) Working principle and experimental demonstration of the dithering technique that allows extracting in real-time the partial derivatives of the MZI transfer function with respect to the electrical power dissipated by the heaters (H1, H2). b) Schematic view of the real-time control scheme applied to a single MZI, allowing automated minimization of the light reaching the photodiodes. c) Seamless extension of the feedback approach to a multi-aperture self-aligning beam coupler, only requiring two dithering signals to control multiple interferometers.

The technique can be straightforwardly extended to control a more complex photonic circuit such as a multi-aperture beam coupler, by implementing multiple identical and independent control loops, as shown in Figure S1c for the case of a binary mesh with 15 MZIs. Thanks to the particular structure of the photonic circuit, requiring sequential configuration of each stage of the mesh, all the control loops can exploit the same dithering signals [37]. Indeed, once the MZIs of a certain stage of the receiver are successfully configured, the residual dithering oscillation at their output is zeroed and does not affect the configuration of the following devices. The interaction between stages only happens during an initial transitory phase and it does not impair the correct operation of the mesh. This approach simplifies the requirements of the control electronics, since only 2 dithering signals are necessary to control 30 heaters and a single-frequency lock-in readout is sufficient to implement an effective control action on the whole photonic circuit.

Once the mesh configuration is completed, the control loops are kept active to counteract dynamic perturbation effects in real-time, such as turbulence-induced amplitude and phase fluctuations of the incoming optical beam and thermal variations. Interestingly, the bandwidth (BW) of the control system is defined by the response time of the single MZI (time constant of 400 µs, corresponding to a BW around 400 Hz in our case) and not of the full receiver. Indeed, if the perturbation is within the bandwidth of the single MZI control, each feedback loop is able to constantly update the heaters voltage and keep the interferometers in the correct operating point, thus avoiding interactions between the stages of the mesh. Therefore, the chip can successfully counteract the effect of realistic turbulence effects, as shown in the following experiments.

## Supplementary section 2: experimental free-space setup

The experimental setup employed in the experiments presented in this work is shown in Figure S2. It has been conceived to emulate an outdoor medium-distance link in an indoor controlled environment, thus easing system-level testing.

The laser source (wavelength centered at 1550 nm) and the optical modulator are integrated into the ID Photonics IQ Optical Multi-Format Transmitter. The 25 Gbaud on-off keying (OOK) and a 4-level pulse amplitude modulation (PAM4) radio-frequency signals are generated using the Multilane ML4039EN BERT. The resulting optical signal is amplified by an erbium-doped fiber amplifier (EDFA, IPG Photonics EAD-1K-C) up to a power level of 25 dBm before being coupled to a fiber collimator. Such an EDFA is used to compensate for the insertion loss of the modulator and the coupling losses between the FSO beam and the input apertures (2D array of grating couplers) and at the chip output port. The amplified spontaneous emission (ASE) noise of the EDFA is reduced with an ASE rejection filter, with a −3 dB bandwidth of 0.25 nm centered around the carrier wavelength. A fiber polarization controller is used to align the polarization state of the light at the chip input to the transverse electric (TE) mode of the grating coupler.

The FSO beam is generated with a fiber collimator coupled to a free-space beam expander, made of a biconcave lens (LD2060-C) and a biconvex lens (LB1630-C) with a maximization of 6.6. In this way, a free-space beam with a waist of around 5 cm is generated after a free-space propagation of 3 meters. A folded setup, where the light beam is reflected with a spatial light modulator (SLM) after half its propagation, has been employed to reduce the area occupation of the setup. The SLM can also be used to steer the beam between the chip input and a near-infrared camera (Xenics Bobcat 640), which is useful for monitoring and characterizing the effect of turbulence. A heat gun is included in the setup for generating dynamic perturbations (in this configuration, the SLM works as a plane mirror and does not introduce any perturbation of the phase profile of the beamfront). By changing the SLM mask or gun position, we could vary the turbulence strength from an effective refractive index parameter $C_n^2$ of $10^{-14}$ $m^{-2/3}$ up to $10^{-10}$ $m^{-2/3}$. The results shown in this section have been obtained by emulating a turbulence strength in the order of $10^{-12}$ $m^{-2/3}$. Due to the size of the beam and the short length of the link, the wander effect was negligible.

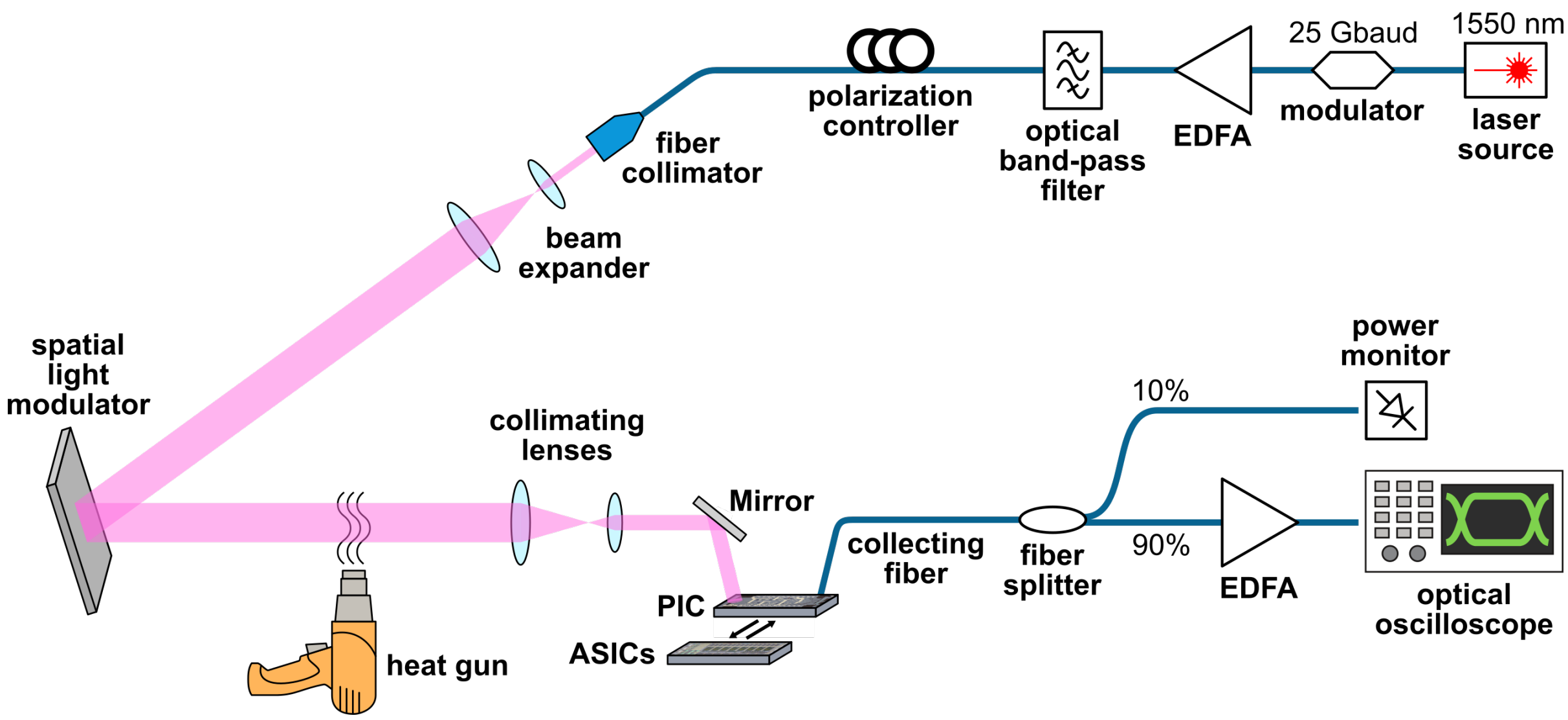

**Fig. S2 Experimental setup.** Schematic view of the free-space experimental setup for testing the integrated circuits.

A second set of lenses (LB1889-C f=250 and LD2060-C f=-15) is placed before the chip input to shrink the beam size down to the dimension of the integrated optical antenna array (diameter of around 360 μm). A turning mirror is also positioned on top of the photonic chip to steer the horizontal optical beam to the tilt angle of the on-chip grating couplers (12° with respect to the normal of the chip surface). A single-mode (SM) fiber is positioned on top of the output grating coupler using a 3-axis micro-positioner. A fiber splitter is then used to direct 10% of the optical power towards a bench-top monitor photodetector (HP 81521B), used to assess the system state and characterize the turbulence temporal dynamics. The rest of the light is sent to an EDFA (Amionics EDFA-PA-35-B-FA) to amplify the signal to match the dynamic range of the high-speed optical oscilloscope (Tektronix, DPO75002SX) to evaluate the transmission performance.

The photonic and electronic chips are mounted on a custom-printed circuit board (PCB). The board allows easy electrical routing to the integrated circuits. It provides the power supply to the ASICs, as well as the reference voltages necessary for their operations. It is also needed to interface the chips with a personal computer, for real-time monitoring of the system state. The shape of the PCB has been conceived to allow easy optical access to the photonic chip, both at the free-space input side and at its output with an optical fiber. The temperature of the assembly is monitored with a thermistor placed close to the photonic chip and stabilized to 28°C using a thermo-electric cooler (TEC), mounted underneath the PCB.

## Supplementary section 3: transmission performance with static turbulence

The spatial light modulator (SLM) has been used to introduce static turbulence in the FSO link with a well-defined profile. The phase screens have been properly engineered to simulate free-space propagation for a distance of a few hundred meters, with a perturbation having an effective refractive index parameter $C_n^2$ of $10^{-12}$ $m^{-2/3}$. Figure S3 reports the 15 different phase masks employed in the experiment, where the electronic-assisted universal beam coupler (UBC) has been used to compensate for the effect of the perturbation. Figure 5c in the main manuscript shows the collected optical power at the chip output, when the ASICs are activated after changing the phase screens and when a static UBC configuration is maintained for all the masks. As expected, a signal with higher average optical power and lower standard deviation is observed when the electronic chips dynamically reconfigure the optical circuit. To further validate this result, we modulated the input light beam with a 25 Gbit/s on-off keying (OOK) non-return-to-zero (NRZ) signal, thus emulating a free-space high-speed communication through a turbulent channel. Figure S3 shows the eye diagrams at the PIC output for each SLM mask, when the control loops are on/off. When the turbulence effect is not counteracted by the adaptive receiver, the transmission quality is completely degraded and closed eye diagrams are observed. Instead, the dynamic reconfiguration of the UBC ensures a correct reception of the modulated signal, as confirmed by the acquired eye diagrams that show a quality factor (QF) with an average value of 5.68 and a standard deviation of 0.6. This result confirms the advantage of the adaptive receiver and certifies the correct control action performed by the electronic chips.

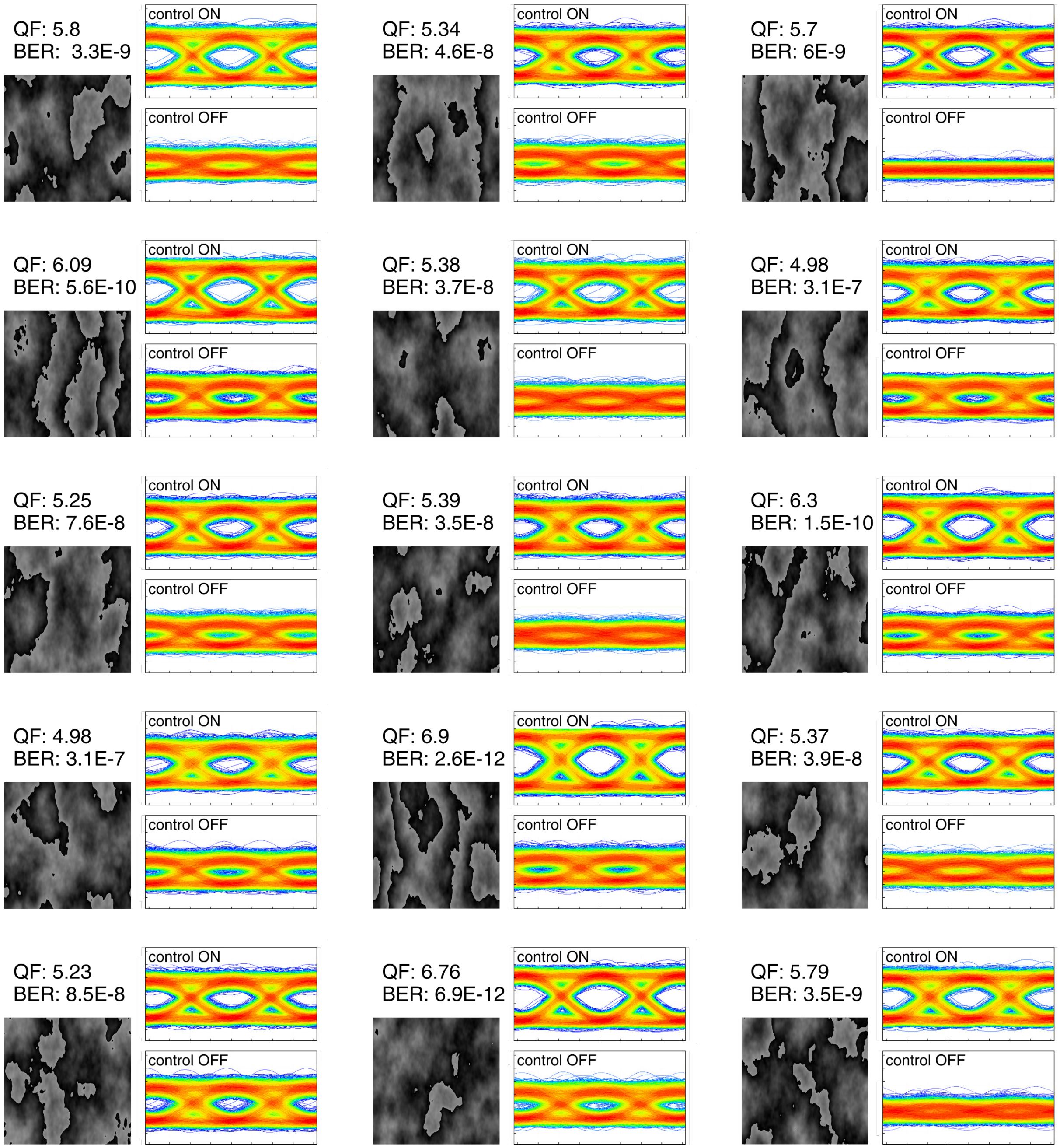

**Fig. S3 Generation and mitigation of static turbulence in the FSO link.** Picture of the 15 SLM phase screens employed to generate a controlled turbulence profile and emulate an outdoor FSO link. The corresponding eye diagrams at 25 Gbit/s at the PIC output when the ASICs are on/off are also reported, demonstrating the successful compensation of the perturbation performed by the chips.

## Supplementary section 4: electronic chip characterization

The electronic application-specific integrated circuit (ASIC) has been electrically tested to verify the performance of both the front-end readout and actuator driving stages. The schematic of a single ASIC channel is reported in Figure S4a for convenience.

The analog front-end, working with a 3.3V power supply, comprises a transimpedance amplifier (TIA), receiving the photodiode (PD) current, a gated integrator (GI), serving as a low-pass filter, and a 10-bit successive approximation analog-to-digital converter (ADC), chosen to limit area and dissipated power. Since the optical power on the detector (responsivity 1 A/W) ranges from 1 mW, when the interferometer is not tuned, to less than 100 nW when the light is steered away from the monitor PD, the front-end has to sense dithering signals over a wide input current range. To cope with the limited resolution of the ADC, the amplifying stage features a variable gain, adjusted automatically according to the sampled value. This solution is implemented by modifying the values of the TIA feedback resistor $R_F$ and the GI resistor $R_G$ to increase the amplification $R_F/R_G$ by 4 each time the input current is scaled of the same factor. Figure S4b shows the measured transfer function of the TIA for 6 values of $R_F$, demonstrating a bandwidth around 1 MHz for whatever configuration, enough to detect dithering signals with a frequency around 10 kHz. The values sampled by the ADC as a function of the input PD current are instead reported in Figure S4c, showing that signals with a dynamic range of more than 50 dB can be correctly acquired by the converter thanks to the adaptive amplification stage.

The actuator driving circuit comprises a 12-bit digital-to-analog converter (DAC) and a high-current driver. The resolution of the DAC has been chosen to provide an accuracy of about 1.5 mV on the heaters control voltage. This corresponds to an uncertainty of the phase shift of about 2 mrad in the worst case, which, in an ideal MZI, would lead to negligible crosstalk of about −60 dB. The DAC has been designed with a monotonic transfer function, which guarantees that the sign of the feedback loop is not accidentally inverted during the operation of the ASIC. The DAC characterization in terms of differential and integral non-linearity (DNL and INL, respectively), expressed as fraction of least-significant bit (LSB), is reported in Figure S4d, confirming the monotonic behavior of the converter (DNL always below 1 LSB). The high-current driver is instead needed for providing up to 15 mA to the thermal actuators, corresponding to around 4π of phase shift in the MZI transfer function. This wide operating range is required to always ensure convergence of the feedback algorithm, regardless of the initial heater voltage. Being the generated phase shift proportional to the power dissipated by the actuators, a quadratic nonlinearity is inherently present in the transfer function from the control voltage to the output optical power. In order to linearize the control loop, an approximated square root digital operation is performed on the digital value fed to the DAC. Such a solution trades precision for compactness, since it occupies only 4% of the area of the digital chip area. Figure S4e shows the current provided to the actuators by the driving stage, and the corresponding heater dissipated power, confirming the effective linearization performed by the square root operator. The same result can be observed in Figure S4f, which reports the optical power at the output of a MZI as a function of the internal heater voltage, when the square root operation is activated/deactivated. As expected, a more linear phase shift is generated by the chip when the digital compression is activated.

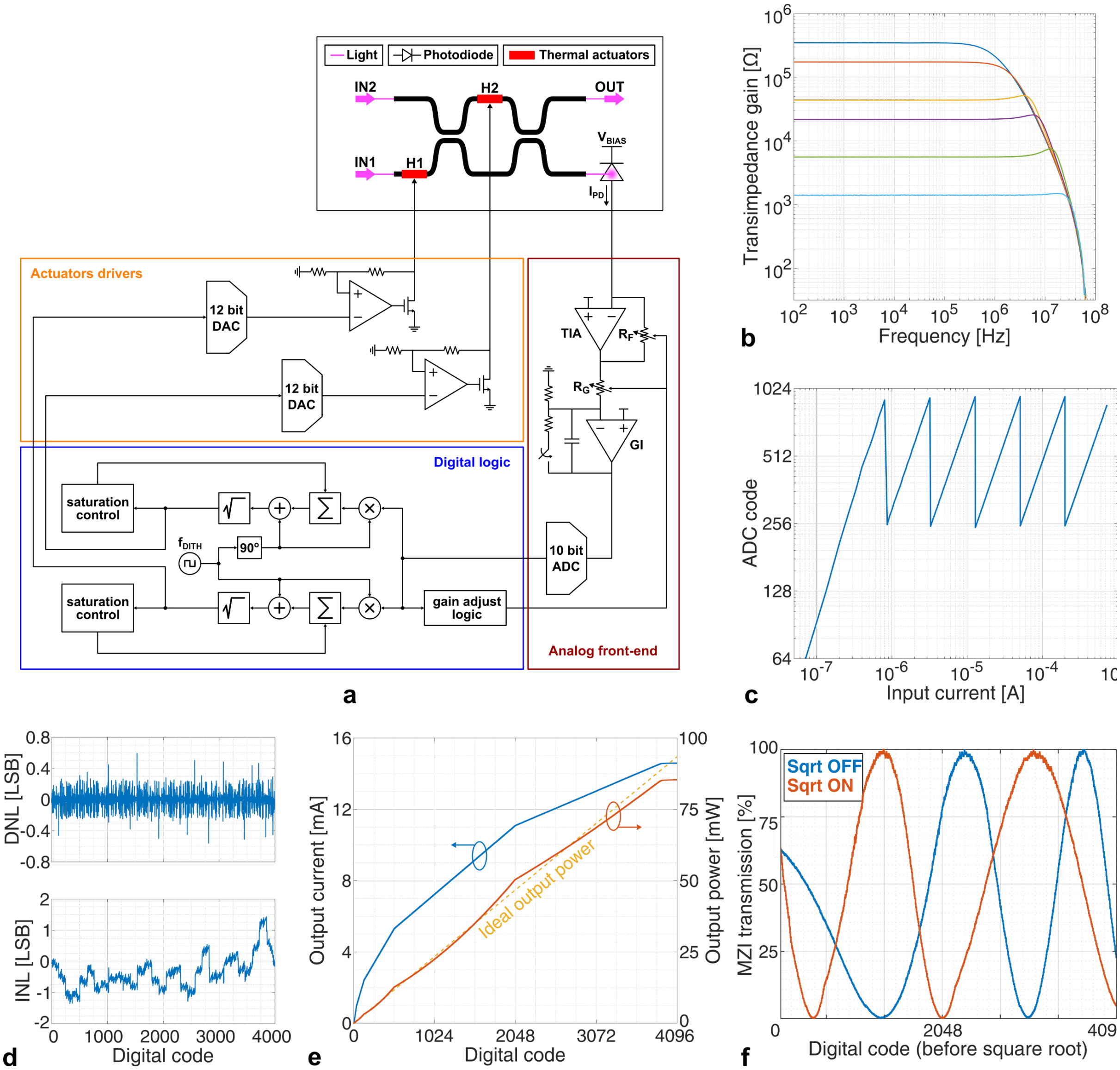

**Fig. S4 Electrical characterization of the ASIC.** a) Schematic view of the architecture of a single ASIC channel. b) Frequency response of the TIA for different gain values. c) Automatic adjusting of the scale of the ADC over the full range of input current. d) Characterization of the DAC non-linearities and e) validation of the square root approximation. f) MZI transfer function obtained by sweeping the heater voltage, with square root compression on and off.

## Supplementary section 5: photonic building blocks and PIC loss analysis

The self-aligning beam coupler has been designed in standard Silicon Photonics technology. The chip has been manufactured in an active multi-project wafer run by Advanced Micro Foundry, Singapore [44]. Figure S5 summarizes the main photonic building blocks that have been employed in the design of the circuit. The footprint and the main distinctive features of each device are also reported.

The optical antenna array (OOA) has been designed to maximize the received power in presence of turbulence in the free-space link. To do so, the size of the apertures must be comparable to the coherence radius $r_0$ of the received turbulent beam, they must be separated by at least $r_0$, and must be as many as possible to minimize geometrical losses [45]. In our case, the OAA has been designed based on simulation results for a scintillation index $\sigma_I^2 = 0.01$ and $r_0 = 4$ cm, corresponding to an outdoor link several hundred meters long [46, 47].

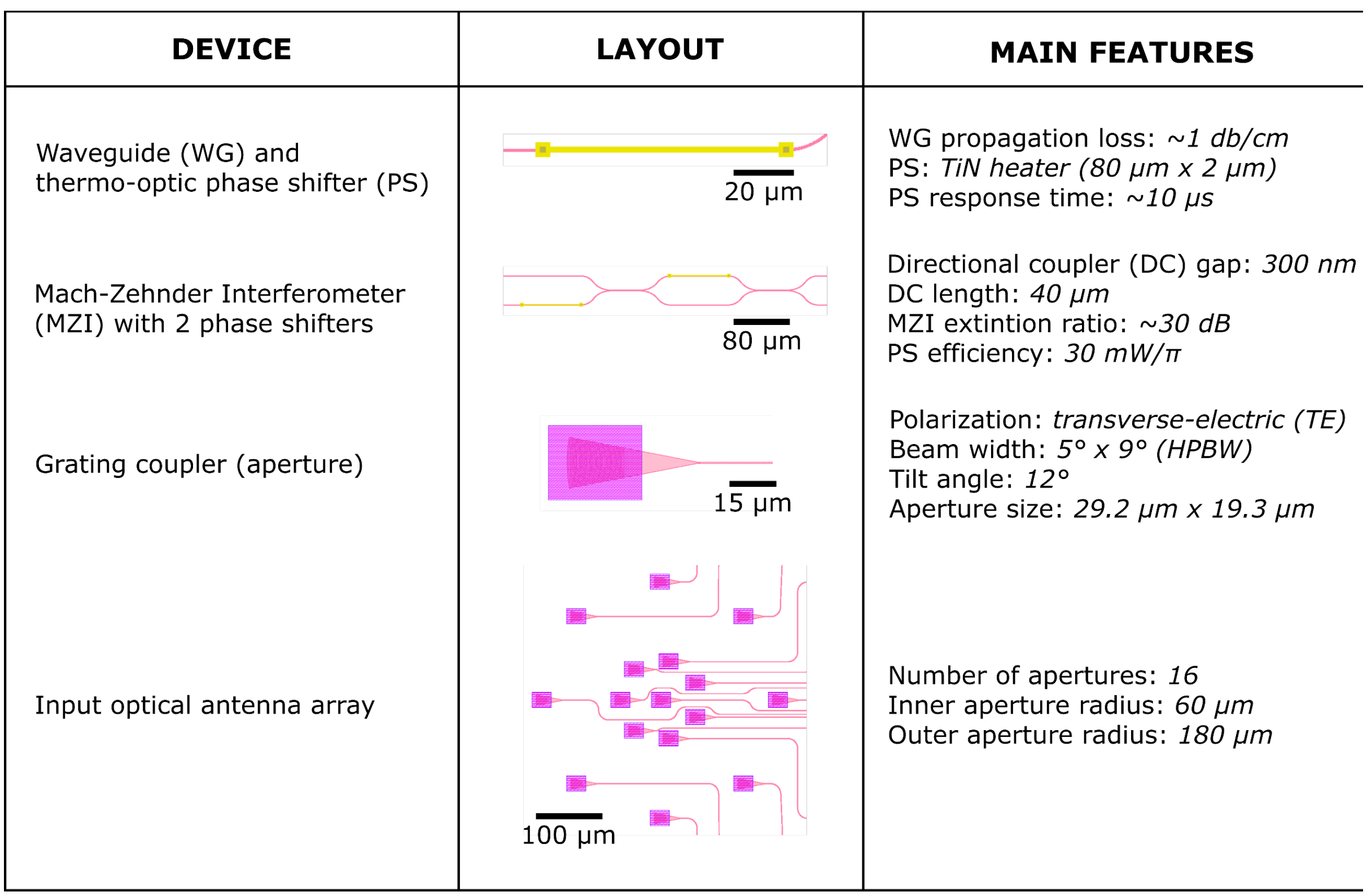

| DEVICE | LAYOUT | MAIN FEATURES |
|---|---|---|
| Waveguide (WG) and thermo-optic phase shifter (PS) | 20 µm | WG propagation loss: *~1 db/cm*<br>PS: *TiN heater (80 µm x 2 µm)*<br>PS response time: *~10 µs* |
| Mach-Zehnder Interferometer (MZI) with 2 phase shifters | 80 µm | Directional coupler (DC) gap: *300 nm*<br>DC length: *40 µm*<br>MZI extintion ratio: *~30 dB*<br>PS efficiency: *30 mW/π* |
| Grating coupler (aperture) | 15 µm | Polarization: *transverse-electric (TE)*<br>Beam width: *5° x 9° (HPBW)*<br>Tilt angle: *12°*<br>Aperture size: *29.2 µm x 19.3 µm* |
| Input optical antenna array | 100 µm | Number of apertures: *16*<br>Inner aperture radius: *60 µm*<br>Outer aperture radius: *180 µm* |

**Fig. S5 Design of the optical beam coupler.** Main building blocks used to implement the integrated photonic processors and their relevant characteristics.

The loss breakdown along the entire optical system is computed as follows. Each grating coupler has about 4.5 dB coupling loss when coupled with standard optical fibers, which translates to about 11 dB insertion loss for a fiber-waveguide-fiber coupling (on-chip loss of the photonic processor is around 2 dB). When the free-space beam generated by the fiber collimator is coupled to the 2D array of the photonic processor, a geometrical loss of about 20 dB is added due to the limited fill factor of the antenna array. This loss can be effectively reduced by improving the fill factor using an array of lenslets at the chip input. To this end, 3D printing techniques, such as two-photon polymerization (TPP), can be used to build custom-designed free-form optical elements directly on top of the photonic chip, thus facilitating optical alignment and reducing assembly and packaging costs [48]. An additional 2 dB loss is due to the aberration of the optical system and possible minor alignment tolerances in the setup. Therefore, the end-to-end loss (from fiber collimator to fiber coupling with output waveguide) is about 30 dB.

## Supplementary references